# CONTENT MANAGEMENT IN RUBY ON RAILS

Antonio Tapiador, Joaquín Salvachúa
*Universidad Politécnica de Madrid*
*Avda. Complutense 30, Madrid, Spain.*

**ABSTRACT**

Web development is currently driven by model-view-controller (MVC) frameworks. How has content management adapted to this scenario? This paper reviews content management features in Ruby on Rails framework and its most popular plug-ins. These features are distributed among the different layers of the MVC architecture.

## 1. INTRODUCTION

How has content management adapted to the arrival of web development frameworks? Using frameworks for web development has become a common practice. Model-view-controller (MVC) patterns facilitate development. They hide complexity, give structure and consistence and promote best practices. Their code is better tested. Finally, a framework becomes popular if it has something useful to offer (Johnson 2005).

On the other hand, content management is the process behind matching what your organization has to what your audience wants (Boiko 2001). It comprises collection, management and publishing content to any outlet. Web content management is the result of delivering content to the web. Web content management became popular with the growth of web pages (McKeever 2003). But, did it catch up with the emergence of web development frameworks?

We have hardly found related work in literature filling the gap between MVC and content management.
There is recent work that introduces the implementation of a web content management system using J2EE MVC technologies (Liduo et al. 2010). It presents a successful case using a 3-tier (MVC) architecture and collects the requisites for web content management. However, it does not explain how content management features are supported by a MVC framework.

In this article, we explore this issue from our experience building web content management systems with Ruby on Rails, a popular web development framework developed to increase productivity. It implements MVC architecture, and relies on "convention over configuration" and "don't repeat yourself" (Bachle and Kirchberg 2007).

## 2. METHOD

We have wide experience developing Ruby on Rails applications, including the GlobalPlaza (http://globalplaza.org/), a Web content management system developed in the context of the EU 7[th] FP Global project.
We have reviewed the implementation of content management features (McKeever 2003, Liduo 2010) in Ruby on Rails and its plug-ins. Table 1 shows each feature and where it is implemented: Rails or a external plug-in. We have measured their popularity relative to Rails (which is the most popular project). The most popular full-featured content management project built with Ruby on Rails (RadiantCMS) is also included.

Popularity is measured using The Ruby Toolbox (http://ruby-toolbox.com/), a web site that collects projects from Github. Github (http://github.com/) is the web site where the Rails community lives in. The score of each project in The Ruby Toolbox is proportional to the number of developers that are watching it and its forks in Github.

## 3. RESULTS

| Content collection | | Content delivery | |
|---|---|---|---|
| Standard tools for content creation | Rails, Formstastic, WillPaginate | Static content | Rails |
| Multi-user support & authorship | Devise | Dynamic content | Rails |
| Separation of content from presentation | Rails | Automatic link checking | Rails |
| Content syndication | Rails | Data error checking | Rails |
| Content preview | Rails* | Separate environments | Rails |
| Content versioning | VestalVersions | Content version rollback | Vestal versions |
| Relevant content types | Paperclip | Multi-channel support | Rails, Prawn |
| Form support for catalogue type data | Rails | Automatic site changes | Rails* |
| Localization | Rails | Content personalization | Devise |
| Shared database for content storage | Rails | Control and administration | |
| Thin client | Rails | Role definition and user security | Devise, CanCan |
| Real time access to CM functions | Rails | Taxonomy | ActsAsTaggableOn |
| Workflow | | Audit trail | Rails* |
| Flexible, multi-threaded | AASM | Reporting functions | Rails*, Devise |
| Workflow monitoring and control features | Rails | | |
| Workgroups | CanCan | Rails* means that feature needs implementation | |

**Table 1: Content management features (McKeever 2003, Liduo 2010)
and their support by Rub on Rails or an external plug-in**

## 4. DISCUSSION

Almost all the content management features are available in the Ruby on Rails development framework. Most of them are integrated in the core framework, but some of them are available as plug-ins.

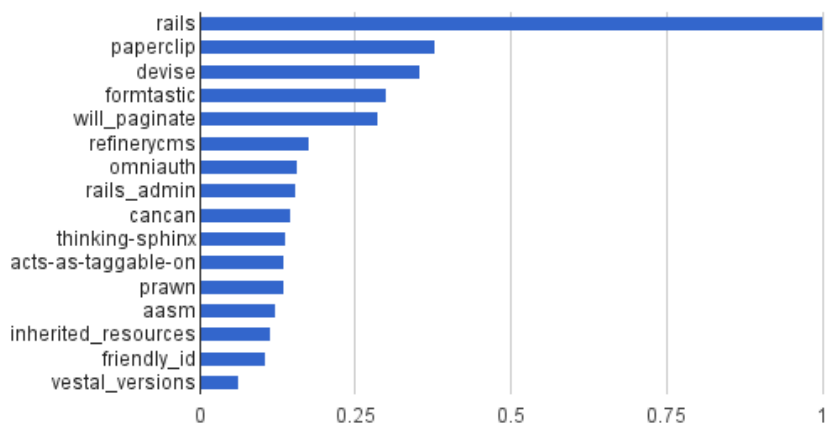

**Figure 1 Popularity of Ruby on Rails content management plug-ins**

Web **authentication** means have evolved a lot in the last years. The most popular authentication plug-in (*Devise*) provides methods like **User name and password**: nowadays it is implemented by almost every web site providing authentication. Credentials can be provided as parameters of a POST request, as result of filling a web form, or through HTTP authentication headers such as basic or digest authentication (Franks 1999). **Access token**: A token is generated and stored in the server associated to the user. This token is passed as a parameter in the request URL, or stored. in the user client, as a cookie. This method is useful for syndication feeds, or remembering user authentication in the browser of a trusted computer. **OpenID** (Recordon and Reed 2006). The user-centric framework emerged as a solution to the "multiple user name and password" problem. OpenID's aims is managing only the authentication of you identity provider. The rest of web sites rely authentication on it. **OAuth**: Initially a protocol for secure API authorization, OAuth (Hammer-Lahav 2010) has become a popular authentication mean used by Facebook and Twitter, among others. Other features like confirm email address, recover passwords, track user information like IP address, timeout session or lock user account. *Devise*'s authentication methods are configurable in the model, letting the content management developer decide which methods are appropriate for each case. It provides with custom controller and views for authentication mechanisms. Besides, it provides with helping methods so the developer can check if the user is authenticated and which is his identity.

**Authorization** is transverse to the MVC architecture. **Model**: the state of the data in the persistence layer (e.g. roles assignations and resources relationships) determines whether authorization is granted or denied. **Controller**: authorization mandate which actions can be performed in the business logic layer. **View**: the interface changes depending on authorization issues. For example, some links are displayed if the user has rights to perform the actions behind them. *CanCan*, provides methods for controller and views. The authorization policies are decoupled from the MVC architecture. They are described in a separate *Ability* class, which it is instantiated for the user in every request.

Ruby on Rails follows **resource oriented architecture (ROA)**. The framework provides resources management support at the three levels of the MVC architecture. Resources are tight related with the life cycle of content, collecting, managing and publishing (Boiko 2001). At the model level, AASM state declarations for workflows, VestalVersions revisions and FriendlyID slug generation. At the controller, inherited resources provide the basic functionality for the life cycle of resources. Several plug-ins enhance the views. Formstatic powering forms generation, WillPaginate for index paginations and Prawn for PDF views. Finally, there are transversal plug-ins to the MVC architecture. These provide functionality at several levels of the MVC architecture. Examples are Paperclip for file management. RailsAdmin interface for resources management and ActsAsTaggableOn for taxonomies and folksonomies.

## 5. CONCLUSION

Content management features keep up with Ruby on Rails development framework. Most of them are integrated in the framework itself, while others are distributed as plug-ins. However, some of them are more popular than others, even than the most popular full-featured content management solution. Content management features are transversal to the MVC architecture, they use some or even all the layers.